\documentstyle[prl,aps,twocolumn]{revtex}
\begin{document} 
\draft 
\title{Uniform semiclassical approximations for umbilic bifurcation
catastrophes}
\author{J\"org Main and G\"unter Wunner}
\address{Institut f\"ur Theoretische Physik I,
Ruhr-Universit\"at Bochum, D-44780 Bochum, Germany}
\maketitle

\begin{abstract}
Gutzwiller's trace formula for the semiclassical density of states
diverges at the bifurcation points of periodic orbits and has to be
replaced with uniform semiclassical approximations.
We present a method to derive these expressions from the standard 
representations of the elementary catastrophes and to directly relate 
the uniform solutions to classical periodic orbit parameters,
thereby circumventing the numerical application of normal form theory.
The technique allows an easy handling of ungeneric bifurcations with 
corank 2 such as the umbilic catastrophes and is demonstrated on a 
hyperbolic umbilic in the diamagnetic Kepler problem.
\end{abstract}


Gutzwiller's periodic orbit theory \cite{Gut67,Gut90} has become the key for
the semiclassical interpretation of quantum systems with underlying
chaotic classical dynamics.
The contributions of isolated periodic orbits to the periodic orbit sum
are given as
\begin{equation}
 {\cal A}_{\rm po} = {T_{\rm po}e^{i(S_{\rm po}/\hbar-{\pi\over2}\mu_{\rm po})}
  \over\sqrt{|\det(M_{\rm po}-I)|}}
\label{A_po}
\end{equation}
with $T_{\rm po}$, $S_{\rm po}$, $M_{\rm po}$, and $\mu_{\rm po}$
the orbital period, classical action, stability matrix, and Maslov index,
respectively.
However, Eq.\ \ref{A_po} diverges at the bifurcation points of periodic 
orbits where orbits are not isolated, and (\ref{A_po}) must be replaced 
with uniform approximations given in terms of diffraction catastrophe 
integrals \cite{Alm87,Sie96}.
The study of bifurcations and uniform approximations is of
fundamental importance to the complete understanding and semiclassical
treatment of systems with mixed regular-chaotic classical dynamics 
\cite{Scho97}.
The derivation of the uniform solutions based on a canonical transformation 
of the coordinates and momenta to normal form coordinates \cite{Mey70,Sad95}
and the construction of the diffraction integrals in terms of the new 
variables is usually a lengthy and nontrivial task 
\cite{Alm87,Sie96,Scho97,Mai97}, 
especially in the neighborhood of bifurcations of codimension $K \ge 2$,
and for catastrophes of corank 2 such as the umbilics.
A simple scheme would be desirable to construct
uniform approximations from classical periodic orbits and to relate
the parameters of catastrophe diffraction integrals directly to the
periodic orbit parameters, such as the classical action $S$ and the 
eigenvalues of the stability matrix $M$.

\begin{figure}[t]
\vspace{8.8cm}
\includegraphics{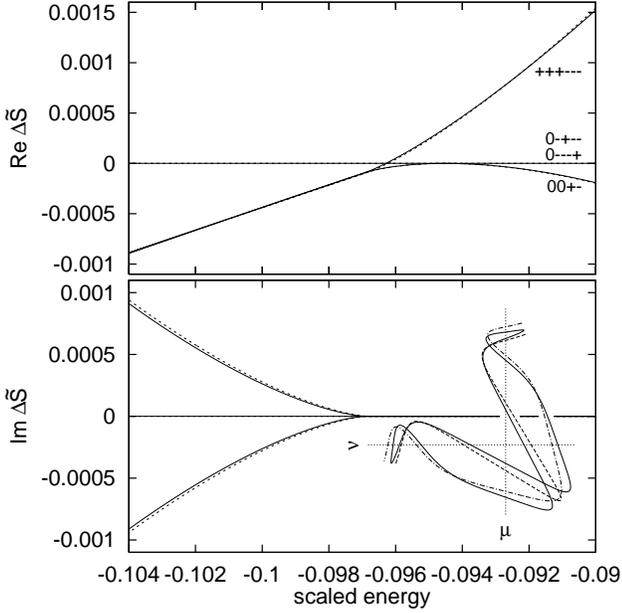}
\caption{\label{fig1} 
Difference $\Delta\tilde S$ between the classical action of the 
four periodic orbits involved in the bifurcations.
Dashed lines: Analytical fits related to the hyperbolic umbilic catastrophe.
Inset: Graphs of periodic orbits {\tt 0-+--} and its time reversal {\tt 0---+}
(solid line), {\tt 00+-} (dashed line), and {\tt +++---} (dashed-dotted line)
drawn in semiparabolical coordinates $\mu=(r+z)^{1/2}$, $\nu=(r-z)^{1/2}$.
}
\end{figure}

\begin{figure}[t]
\vspace{8.8cm}
\includegraphics{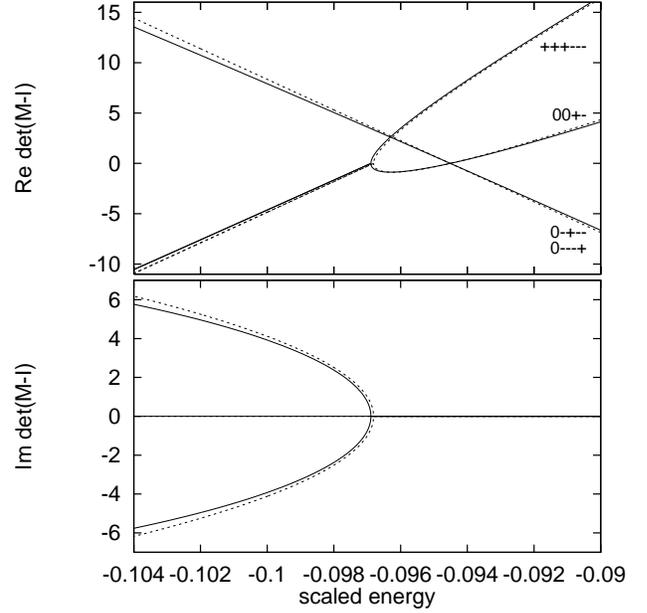}
\caption{\label{fig2} 
Same as Fig.\ 1 but for the determinant $\det(M-I)$ of the periodic orbits.
}
\end{figure}

In this Letter we want to demonstrate that in practical applications the 
derivation of uniform semiclassical approximations can be considerably 
simplified especially for ungeneric bifurcations of codimension $K \ge 2$ 
and catastrophes of corank 2 when starting directly from the standard 
representation of the elementary catastrophes \cite{Pos78,Ber80}.
We illustrate our method by way of example of the diamagnetic Kepler problem
\cite{Fri89,Has89} given by the Hamiltonian 
[in atomic units, $\gamma=B/(2.35\times 10^5~{\rm T})$, and $L_z=m\hbar=0$],
\begin{equation}
 H = {1\over 2}{\bf p}^2 - {1\over r} + {1\over 8} \gamma^2 \rho^2 \; ,
\label{Ham_fkt}
\end{equation}
which is a scaling system, with $w=1/\hbar_{\rm eff}\equiv\gamma^{-1/3}$ 
the scaling parameter and $\tilde E \equiv E\gamma^{-2/3}$ the scaled energy.
The classical dynamics is determined by the scaled energy $\tilde E$ but
does not depend on $w$.
We investigate the birth of four periodic orbits through two close-by
bifurcations near the scaled energy $\tilde E \approx -0.096$ where we
search for both real and complex ``ghost'' orbits \cite{Mai97,Kus93}.
For the nomenclature of the real orbits we adopt the symbolic code of Ref.\
\cite{Eck90}.
At scaled energy $\tilde E_b^{(1)}=-0.09689$, the two orbits {\tt 00+-} and 
{\tt +++---} are born in a tangent bifurcation.
At energies $\tilde E<\tilde E_b^{(1)}$, a prebifurcation ghost orbit and
its complex conjugate exist in the complex continuation of the phase space.
Orbit {\tt 00+-} is born unstable, and turns stable at the slightly higher
energy $\tilde E_b^{(2)} = -0.09451$.
This is the bifurcation point of two additional orbits, {\tt 0-+--} and its
time reversal {\tt 0---+}, which also have ghost orbits as predecessors.
The graphs of the real orbits at energy $E=0$ are shown as insets in Fig.\ 
\ref{fig1},
and the classical periodic orbit parameters are presented as solid lines in 
Figs.\ \ref{fig1} and \ref{fig2}.
Fig.\ \ref{fig1} shows the difference in scaled action, 
$\Delta\tilde S \equiv \Delta S/(2\pi w)$, between the orbits.
The action of orbit {\tt 0-+--} (or its time reversal {\tt 0---+}), which is 
real also for its prebifurcation ghost orbits, has been taken as the 
reference action.

At this point the usual procedure would be to investigate the classical
dynamics in the neighborhood of the periodic orbits by numerical application
of normal form theory \cite{Mey70,Sad95}.
The representation of the dynamics in normal form coordinates would finally
lead to the correct type of the catastrophe diffraction integral related to 
the uniform semiclassical approximation with numerically well determined 
coefficients.
However, the numerical procedure of local canonical transformations to
normal form coordinates, e.g., by means of local Fourier-Taylor series 
expansions with numerically obtained coefficients \cite{Sad95} is rather 
lengthy and tedious especially for bifurcations related to catastrophes of 
higher codimension or corank.
The main result of this Letter is to demonstrate that there is a shortcut
to the usual procedure which allows 
\newpage
\phantom{}
\noindent
circumventing the numerical application 
of normal form theory.
By our new method, an easy construction of uniform semiclassical 
approximations for ungeneric types of catastrophes, e.g.\ the umbilics, 
becomes feasible for the first time.

Choosing the elementary catastrophe diffraction integrals as the ansatz
for the uniform semiclassical approximation, we must be able to identify
the stationary points of the exponents with the periodic orbits, i.e., in 
our example four stationary points must exist.
From the seven ``elementary catastrophes'' of Refs.\ \cite{Pos78,Ber80}
this is the case only for the swallowtail and the elliptic and hyperbolic
umbilic.
The correct choice in our example turns out to be the hyperbolic umbilic 
catastrophe, which is of importance, e.g., for uniform $S$ matrix 
approximations in semiclassical scattering theory \cite{Uze82}.
It is a corank 2 catastrophe, i.e., the diffraction integral is 
two-dimensional,
\begin{equation}
 \Psi(x,y) = \int_{-\infty}^{+\infty}dp \int_{-\infty}^{+\infty}dq
   e^{i\Phi(p,q;x,y)}
\label{Psi_def}
\end{equation}
with
\begin{equation}
 \Phi(p,q;x,y) = p^3 + q^3 + y(p+q)^2 + x(p+q) \; .
\end{equation}
\newpage
\phantom{}
\noindent
For our convenience the function $\Phi(p,q;x,y)$ slightly differs from
the standard polynomial of the hyperbolic umbilic given in Ref.\ \cite{Ber80}
but the diffraction integral (\ref{Psi_def}) can be easily transformed to the
standard representation.
The four stationary points of the integral (\ref{Psi_def}) are readily 
obtained from the condition $\nabla\Phi=0$ as
\begin{equation}
 p_0 = -q_0 = \pm\sqrt{-x/3}
 \Rightarrow \Phi(p_0,q_0;x,y) = 0
\label{stat_points_a}
\end{equation}
and
\begin{eqnarray}
 p_0 = q_0 &=& -{2\over 3}y \pm\sqrt{{4\over 9}y^2-{x\over 3}} \Rightarrow
   \nonumber \\
 \Phi(p_0,q_0;x,y) &=& {4\over 3}y\left({8\over 9}y^2-x\right)
   \mp 4\left({4\over 9}y^2-{x\over 3}\right)^{3/2} \; .
\label{stat_points_b}
\end{eqnarray}
The function $\Phi(p_0,q_0;x,y)$ must now be adapted to the classical action 
of the four periodic orbits, i.e., 
$\Delta S = 2\pi w\Delta\tilde S \approx \Phi(p_0,q_0;x,y)$,
which is well fulfilled for
\begin{equation}
 x = a w^{2/3} \left(\tilde E - \tilde E_b^{(2)}\right) \; ; \quad
 y = b w^{1/3} \; ,
\label{xy_def}
\end{equation}
and constants $a=-5.415$, $b=0.09665$, as can be seen from the dashed lines 
in Fig.\ \ref{fig1}.
Note that the agreement holds for both the real and complex ghost orbits.

The next step to obtain the uniform solution is to calculate the diffraction 
integral (\ref{Psi_def}) within the stationary phase approximation.
For $\tilde E > \tilde E_b^{(2)}$ there are four real stationary points
$(p_0,q_0)$ (see Eqs.\ \ref{stat_points_a} and \ref{stat_points_b}), and 
after expanding $\Phi(p,q;x,y)$ around the stationary points up to second 
order in $p$ and $q$, the diffraction integral becomes the sum of
Fresnel integrals, viz.\
\begin{equation}
 \Psi(x,y) \stackrel{x\ll 0}{\sim} {2\pi\over\sqrt{-3x}} + \sum_{+,-}
 {\pi e^{i\left[{4\over 3}y\left({8\over 9}y^2-x\right)\mp
 4\left({4\over 9}y^2-{x\over 3}\right)^{3/2}\pm{\pi\over 2}\right]} \over
 \sqrt{(4y^2-3x)\mp 2y\sqrt{4y^2-3x}}} \, .
\label{Psi_asym}
\end{equation}
The terms of Eq.\ \ref{Psi_asym} can now be compared to the standard
periodic orbit contributions (\ref{A_po}) of Gutzwiller's trace formula.
In our example the first term is related to the orbit {\tt 0-+--} (with a
multiplicity factor of 2 for its time reversal {\tt 0---+}), and the other 
two terms are related to the orbits {\tt 00+-} and {\tt +++---} for the upper 
and lower sign, respectively.
The phase shift in the numerators describe the differences of the action 
$\Delta S$ and of the Maslov index $\Delta\mu=\mp 1$ relative to the 
reference orbit {\tt 0-+--}.
The denominators are, up to a factor $cw^{1/3}$, with $c=0.1034$, the 
square root of $|\det(M-I)|$, with $M$ the stability matrix.
Fig.\ \ref{fig2} presents the comparison for the determinants obtained from 
classical periodic orbit calculations (solid lines) and from Eqs.\
\ref{xy_def} and \ref{Psi_asym} (dashed lines).
The agreement is very good for both the real and complex ghost orbits,
similar to the agreement found for $\Delta\tilde S$ in Fig.\ \ref{fig1}.
The constant $c$ introduced above determines the normalization of the
uniform semiclassical approximation for the hyperbolic umbilic bifurcation
which is finally obtained as
\begin{eqnarray}
 && {\cal A}_{\rm uniform}(\tilde E,w) = (c/\pi) T_0 w^{1/3} \nonumber \\
 && \times \Psi\left(aw^{2/3} (\tilde E - \tilde E_b^{(2)}), bw^{1/3}\right)
 e^{i[2\pi\tilde S_0w-{\pi\over 2}\mu_0]}
\label{A_uniform}
\end{eqnarray}
with $T_0$, $S_0$, and $\mu_0$ denoting the orbital period, action and 
Maslov index of the reference orbit {\tt 0-+--}, and constants $a$, $b$, 
and $c$ as given above.
Note that all parameters are readily determined by classical periodic orbit
calculations.

\begin{figure}[t]
\vspace{9.45cm}
\includegraphics{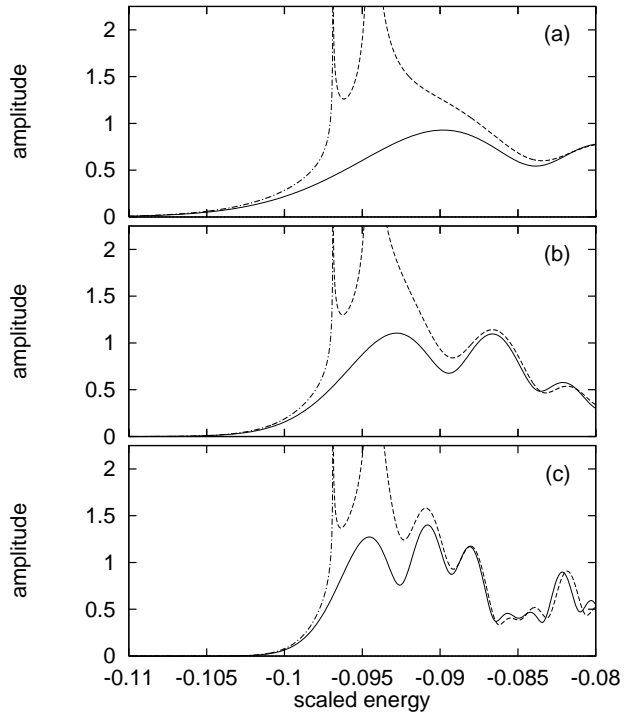}
\caption{\label{fig3} 
Semiclassical amplitudes (absolute values) for magnetic field strength 
(a) $\gamma=10^{-7}$, (b) $\gamma=10^{-8}$, and (c) $\gamma=10^{-9}$
in units of the time period $T_0$.
Dashed line: Amplitudes of the standard semiclassical trace formula.
Dashed-dotted line: Ghost orbit contribution.
Solid line: Uniform approximation of the hyperbolic umbilic catastrophe.
}
\end{figure}

The comparison between the conventional semiclassical trace formula 
(\ref{A_po}) for isolated returning orbits and the uniform approximation 
(\ref{A_uniform}) for the hyperbolic umbilic catastrophe is presented in Fig.\ 
\ref{fig3} at the magnetic field strengths $\gamma=10^{-7}$, $\gamma=10^{-8}$, 
and $\gamma=10^{-9}$, .
For graphical purposes we suppress the highly oscillatory part resulting 
from the function $\exp[i(S_0/\hbar-{\pi\over2}\mu_0)]$ by plotting the 
absolute value of ${\cal A}(\tilde E,w)$ instead of the real part.
The dashed line in Fig.\ \ref{fig3} is the superposition of the isolated 
periodic orbit contributions from the four orbits involved in the bifurcations.
The modulations of the amplitude are caused by the constructive and 
destructive interference of the real orbits at energies 
$\tilde E > \tilde E_b^{(2)}$ and are most pronounced at low magnetic
filed strength (see Fig.\ \ref{fig3}c).
The amplitude diverges at the two bifurcation points.
For the calculation of the uniform approximation (\ref{A_uniform}) we
numerically evaluated the catastrophe diffraction integral (\ref{Psi_def})
using a more simple and direct technique as 
\newpage
\phantom{}
\noindent
described in \cite{Uze83}.
Details of our method which is based on Taylor series expansions will
be given elsewhere \cite{Mai98}.
The solid line in Fig.\ \ref{fig3} is the uniform approximation 
(\ref{A_uniform}).
It does not diverge at the bifurcation points but decreases exponentially 
at energies $\tilde E < \tilde E_b^{(1)}$.
At these energies no real orbits exist and the amplitude in the standard 
formulation would be zero when only real orbits are considered.
However, the exponential tail of the uniform approximation (\ref{A_uniform})
is well reproduced by a ghost orbit \cite{Mai97,Kus93} with positive 
imaginary part of the complex action.
As can be shown, the asymptotic expansion of the diffraction integral
(\ref{Psi_def}) has, for $x\gg 0$, exactly the form of Eq.\ \ref{A_po} but 
with complex action $S$ and determinant $\det(M-I)$ \cite{Mai98}.
The ghost orbit contribution is shown as dashed-dotted line in Fig.\ 
\ref{fig3}.

\begin{figure}[t]
\vspace{4.8cm}
\includegraphics{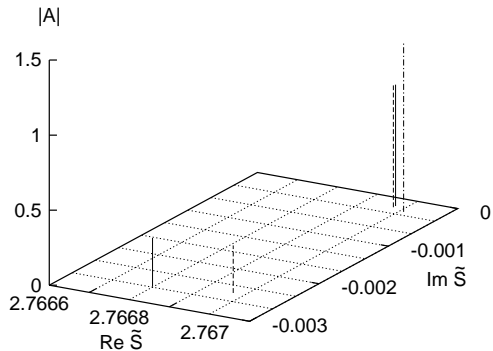}
\caption{\label{fig4} 
High resolution recurrence spectra at scaled energy $\tilde E=-0.1$.
Solid peaks: Part of the quantum recurrence spectra.
Dashed-dotted peak: Classical ghost orbit contribution.
Dashed peaks: Uniform approximation of the hyperbolic umbilic catastrophe.
}
\end{figure}

To verify the hyperbolic umbilic catastrophe in quantum spectra we
diagonalized the Hamiltonian (\ref{Ham_fkt}) in a complete basis set 
(for computational details see, e.g., \cite{Mai94}) at constant scaled energy
$\tilde E=-0.1$, which is slightly below the bifurcation energies, and 
calculated 9715 eigenvalues $w_n$ for the scaling parameter in the region 
$w<140$.
The scaled spectrum was analyzed by the high resolution method of Ref.\
\cite{Mai97a}, i.e., the density of states $\varrho(w)=\sum_n\delta(w-w_n)$ 
was fitted by application of the harmonic inversion technique to the 
functional form of the semiclassical trace formula
\begin{equation}
 \varrho(w) = \sum_k A_k e^{-2\pi i\tilde S_k w}
\end{equation}
with complex parameters $A_k$ and $\tilde S_k$.
For isolated returning orbits these parameters, obtained from the quantum 
spectra, can directly be compared to the periodic orbit parameters of the 
classical calculations \cite{Mai97a}.
The part of the complex action plane which is of interest for the hyperbolic
umbilic catastrophe discussed above is presented in Fig.\ \ref{fig4}.
The two solid peaks mark the positions $\tilde S_k$ and the absolute values 
of amplitudes $|A_k|$ obtained from the quantum spectrum.
However, there is only one classical ghost orbit which is of physical 
relevance (dashed-dotted peak in Fig.\ \ref{fig4}).
The position of that peak is in good agreement with the quantum result
but the amplitude is enhanced, as is expected for isolated periodic 
orbit contributions near bifurcations (see Fig.\ \ref{fig3}).
For comparison we have analyzed the uniform approximation (\ref{A_uniform}) 
at constant scaled energy $\tilde E=-0.1$ and in the same range $0<w<140$ 
by applying the harmonic inversion technique of Ref.\ \cite{Mai97a}.
The results for the uniform approximation are presented as dashed peaks 
in Fig.\ \ref{fig4}.
The two peaks agree well with the quantum results for both the complex
actions and amplitudes.
The enhancement of the ghost orbit peak and the additional non-classical 
peak observed in the quantum spectrum are therefore clearly identified as 
artifacts of the bifurcation, i.e., the hyperbolic umbilic catastrophe.

In conclusion, we have presented a simple method to construct uniform
approximations for the semiclassical density of states and to relate the 
parameters of catastrophe diffraction integrals directly to periodic orbit 
parameters such as classical action and eigenvalues of the stability matrix 
at energies near the bifurcation.
The method is a shortcut to the conventional procedure, i.e., it circumvents
the analysis of classical dynamics in the neighborhood of periodic orbits 
by numerical application of normal form theory, and therefore allows,
for the first time, an easy handling of ungeneric bifurcations of several 
orbits related to catastrophes of higher codimension and corank.
This has been demonstrated by way of example of a hyperbolic umbilic 
catastrophe in the diamagnetic Kepler problem, but evidently the method 
may be applied to other systems and catastrophe types as well.
The technique will be useful for the semiclassical quantization of systems
with mixed regular-chaotic classical dynamics, e.g., in combination with 
the method of harmonic inversion which has been successfully applied to 
systems with complete hyperbolic dynamics \cite{Mai97b}.

\smallskip
We acknowledge stimulating discussions with F.\ Haake, H.\ Schomerus, and 
T.\ Uzer and thank K.\ Wilmesmeyer for assistance with the classical
calculations.
This work was supported by the Deutsche Forschungsgemeinschaft (SFB 237).

\end{document}